\documentclass[final,3p,times,twocolumn]{elsarticle}

\usepackage{amssymb}
\usepackage{amsmath}
\usepackage{bm}

\journal{Chemical Physics Letters}

\begin{document}

\begin{frontmatter}



\title{A screened automated structural search with semiempirical
 methods}

\author[JAEA]{Yukihiro Ota}
\author[JAEA,IMS]{Sergi Ruiz--Barragan}
\author[JAEA]{Masahiko Machida}
\author[JAEA]{Motoyuki Shiga}

\address[JAEA]{CCSE, Japan Atomic Energy Agency, 178-4-4 Wakashiba, Kashiwa
 277-0871, Japan}
\address[IMS]{Department of Theoretical and Computational Molecular
 Science, Institute of Molecular Science, Okazaki 444-8585, Japan}

\begin{abstract}
We developed an interface program between a program suite for an automated
 search of chemical reaction pathways, GRRM, and a program package of
 semiempirical methods, MOPAC. 
A two-step structural search is proposed as an
 application of this interface program. 
A screening test is first performed by semiempirical calculations. 
Subsequently, a reoptimization procedure is done by ab initio or density
 functional calculations. 
We apply this approach to ion adsorption on cellulose. 
The computational efficiency is also shown for a GRRM search.  
The interface program is suitable for the structural
 search of large molecular systems for which semiempirical
 methods are applicable. 
\end{abstract}

%
%
%

\end{frontmatter}


\section{Introduction}
A systematic exploration of chemical reaction pathways is one of the
challenging issues in modern computational chemistry. 
This task is common to many computational calculations, such as the
design of catalytic activity~\cite{Houk;Cheong:2008} 
and the creation of crystal structures from chemical composition
alone~\cite{Woodley;Catlow:2008}. 
Typically, the relevant computational task is to optimize a
multi-dimensional function on potential energy
surfaces~\cite{Jensen:2007} to find all the important local minima and
saddle points. 
The well-ordered manipulation of a large amount of data is
required as well. 
Significant research efforts have devoted to
these topics~\cite{Wales:2003,Lyakhov;Zhu:2013,Maeda;Ohno;Morokuma:2013}.  
A program suite, GRRM~\cite{Maeda;Ohno;Morokuma:2013,GRRM}, is intended
to achieve an automated search of reaction pathways. 
This program suite has been applied to various
issues in chemistry and materials science, such as 
the reaction pathways of oxygen atom on silicon
surfaces~\cite{Ohno;Ohno:2010}, 
exploring conical intersections near the Franck-Condon region in
different molecules~\cite{Maeda;Harabuchi;Morokuma:2014}, 
a bolylation of organic halides with
silyboranes~\cite{Uematsu;Taketsugu:2015}, 
and the prediction of undiscovered carbon
structures~\cite{Ohno;Yamakado:2015}. 

The GRRM program enables an automated search of reaction routes.
However, an efficient search is mandatory for treating large systems, such
as polymers, proteins and biological molecules. 
The search algorithm in GRRM requires the calculation of forces upon the
atoms of a target molecule along the potential energy
surface~\cite{Maeda;Ohno;Morokuma:2013}, calculated by ab inito 
molecular orbital theory or density functional theory (DFT) with
external program packages, such as GAUSSIAN 09~\cite{Gaussian09} and
GAMESS~\cite{Gamess,Gordon;Schmidt:2005}. 
Thus, depending on available computational resource, it is desirable to
reduce the computational effort of force calculations. 

In this article, we describe an interface program between GRRM and
MOPAC~\cite{MOPAC2012,Maia;Rocha:2012} to implement an automated search
of chemical reaction pathways of large molecular system with
semiempirical methods.   
This program is compatible with GRRM 14~\cite{GRRM}. 
The output data of MOPAC is converted into a readable format for input
to GRRM 14, and vice versa. 
Moreover, we propose a two-step procedure to find stable structures in
large molecular systems. 
The first step is a screening test via GRRM with semiempirical methods,
to find the candidates for stable structures. 
The next step is the separate reoptimization of the resultant candidates
with more costly but precise methods either based on ab initio or
density functional calculations. 

The interface program and the two-step structural search are tested 
for ion adsorption on cellulose. 
The mechanism of ion adsorption on materials and organic products is
important for various industrial and environmental issues, such as designing
nanosensors for hydrate fissile ions in waste
water~\cite{Kumar;Seminario:2013}, the applications of graphene oxides
to lithium-ion batteries~\cite{Uthaisa;Barone:2014}, and the transport
mechanism of radioactive cesium ion in plants~\cite{Adams;Shin:2015}. 
We search for the stable structures of cellulose with three
 different alkali cations,
 $\mbox{Li}^{+}$, $\mbox{K}^{+}$, and $\mbox{Cs}^{+}$. 
We also show a way of quantifying the relative stability of cation
binding to molecules, relevant to measurable data in experiments. 
The efficiency of the present approach is discussed from the
 viewpoint of computational performance. 
We show that in a screening test with GRRM 14 the present approach 
 significantly reduces the computational time compared to GAMESS. 
Although semiempirical methods implemented in GAUSSIAN 
 have been used previously with GRRM~\cite{Takayanagi;Tachikawa:2009},
 it is much more efficient to use MOPAC directly, as will be
 evident from the present paper.
Thus, as long as the semiempirical method chosen is qualitatively
 correct, the method proposed herein is highly desirable for large
 molecular systems. 

\section{Methods and computational details}
\label{sec:methods}
Our interface program connecting between GRRM and MOPAC has been
 completed for the combination of GRRM 14 and MOPAC 2012. 
GRRM 14 includes two kinds of main algorithms for searching
 chemical reaction pathways.
One is the anharmonic downward distortion following 
method, leading to an exploration of isomerization and
dissociation pathways. 
The other is the artificial force induced reaction (AFIR) method, leading
 to a search of associative pathways of two or more
 reactants by way of transition states. 
Both methods require forces on the potential energy
surfaces of target systems. 
When calling the interface program, the forces are calculated by
a semiempirical method. 
Then, the standard output data of MOPAC is transformed into that
 readable for GRRM. 
Thus, an automated search of chemical reaction pathways is performed by
a semiempirical method. 

We search for the stable structures of $\alpha$-cellulose 
 $(\mbox{C}_{6}\mbox{H}_{10}\mbox{O}_{5})_{n}$ ($n=1,\,2$) binding
 a single cation, as an application of the present interface program. 
The structure of $\alpha$-cellulose is built up from crystal structure
data~\cite{Nishiyama;Chanzy:2002} at the
B3LYP~\cite{Becke:1993,Stephens;Frisch:1994}/6-31G$\ast\ast$ level of
theory by Gaussian 03. 
The edges of a chain structure of cellulose are terminated by
$\mbox{--OH}$ and $\mbox{--H}$ assuming the product of hydrolysis
 reaction. 
The structure with $n=1$ corresponds to $D$-glucose, but we call it
 cellulose monomer for convenience throughout this article. 
Three kinds of alkali cations, $\mbox{Li}^{+}$, $\mbox{K}^{+}$, and
$\mbox{Cs}^{+}$, are studied. 
A screening test is first performed by the multicomponent AFIR (MC-AFIR)
method~\cite{Maeda;Morokuma:2010,Maeda;Morokuma:2011} with 
PM6~\cite{Stewart:2007} in MOPAC 2012. 
The MC-AFIR method in GRRM 14 allows us to produce different associative
pathways from the randomly-generated configurations of several
reactants, with the aid of multiple artificial forces between intra- and
inter-reactant components. 
We apply the artificial force between the cation and each oxygen atom of 
cellulose, with the upper bound of a collision energy
parameter~\cite{Maeda;Ohno;Morokuma:2013} set to be 
$100\,\mbox{kJ}/\mbox{mol}$. 
The stopping criterion in this search is that an identical reaction
pathway is discovered $50$ times in a row. 
Subsequently, the structures discovered in the above
screening procedure are separately reoptimized at the B3LYP/LanL2DZ
level of theory by GAUSSIAN 09. 
Each structure has an identification number in ascending
order, according to increasing energy. 
The relative stability of structures is assessed by the difference
of calculated energy, $E^{(s)} - E^{(s=1)}$ with the identification
 number of structures $s$ and the calculated energy $E^{(s)}$. 

After the search of stable structures, we study a solvent effect in the
cation binding to cellulose molecules, within the polarizable continuum
model (PCM)~\cite{Tomasi;Persico:1994,Tomasi;Cammi:2005}.
The stable structures in aqueous solutions are calculated at the
B3LYP/LanL2DZ level of theory by DFT calculations with the PCM of water. 
The above gaseous-phase results are utilized as initial structures. 
To study statistical properties in aqueous solutions at a temperature
$T$, we use the canonical ensemble of stable structures. 
We may define the partition function as
\begin{equation}
Z_{\rm PCM}
=
{\textstyle \sum^{\prime}}\, 
{\rm e}^{-E^{(s)}_{\rm PCM}/k_{\rm B}T}, 
\label{eq:partition_function_structures}
\end{equation}
with the Boltzmann constant $k_{\rm B}$. 
The prime symbol on summation means that the
summation index $s$ runs over a set of distinct structures. 
We pick up distinct structures using inter-atomic distance; if all the
inter-atomic distances of the $s$th stable structure are equal to those
of the $s^{\prime}$th structure, the two structures are considered to be
identical. 
The threshold value of distance is set as $0.1\,\mbox{\AA}$. 
The canonical ensemble leads to the probability of finding a certain
structure. 
In this article, we focus on the probability of finding the most stable
structure, 
\(
p_{1} = Z_{\rm PCM}^{-1}{\rm e}^{-E_{\rm PCM}^{(1)}/k_{\rm B}T}
\), at $T=300\,\mbox{K}$. 

The two-step structural search allows us to obtain well-ordered data
of cation-binding patterns, leading to an estimation of experimental
data on ion adsorption, such as the amount of adsorption 
and the rate of ion exchange. 
To address this issue, let us focus on a reaction process 
\begin{equation}
 \mbox{RK}^{+} + \mbox{M}^{+} \to \mbox{RM}^{+} + \mbox{K}^{+}, 
\label{eq:reaction_process}
\end{equation}
where $\mbox{R}$ represents either cellulose monomers or dimers and
$\mbox{M}^{+}$ does a cation. 
Calculating an energy difference between the reactants and the products
would quantify the relative amount of cation adsorption on molecules. 
The energy data of discovered structures is suitable for estimating
this difference. 
In a gaseous phase we can write this quantity as
\begin{eqnarray}
&&
 \Delta E_{\rm gas}
=
[
E^{(s=1)}_{\rm DFT}(\mbox{RM}^{+}) + E_{\rm DFT}(\mbox{K}^{+})
] \nonumber 
\\
&&
\hspace{13mm}
-
[
E^{(s=1)}_{\rm DFT}(\mbox{RK}^{+}) + E_{\rm DFT}(\mbox{M}^{+})
] \label{eq:reaction},
\end{eqnarray} 
with the DFT-calculation energy of an isolated cation, 
$E_{\rm DFT}(\mbox{M}^{+})$. 

Since the experiments of ion adsorption on molecules are typically 
performed in water~\cite{Adams;Shin:2015}, a way of estimating the
energy difference in aqueous solutions, denoted by $\Delta E_{\rm aq}$,
is desirable for linking theoretical and experimental data. 
The two-step structural search with PCM leads to useful data on this
issue. 
On the energy of $\mbox{RM}^{+}$ in
Eq.~(\ref{eq:reaction}), we evaluate the free energy according to
partition function (\ref{eq:partition_function_structures}), 
\(
A_{\rm PCM}
=
-k_{\rm B}T \ln Z_{\rm PCM}
\), at $T= 300\,\mbox{K}$ 
rather than the energy of the most stable structure,  
to take the occurrence of different structures in aqueous solutions into
account. 
The entropy, 
\(
S_{\rm PCM} = - \partial_{T}A_{\rm PCM}
\), is also evaluated, to quantify the number of states in the
statistical distribution of stable structures in aqueous solutions. 
As for an isolated cation, we replace $E_{\rm DFT}(\mbox{M}^{+})$ in
Eq.~(\ref{eq:reaction}) with the energy obtained by DFT calculations
with PCM. 

The efficiency of a GRRM search with MOPAC is examined by measuring the
computational time for discovering the stable structures of cation
binding to cellulose compared to GAMESS. 
Moreover, measuring a force-calculation time of DFT calculations, we
estimate the computational time of a GRRM structural search with DFT. 

\begin{figure}[tbp]
\centering
\scalebox{0.196}[0.196]{\includegraphics{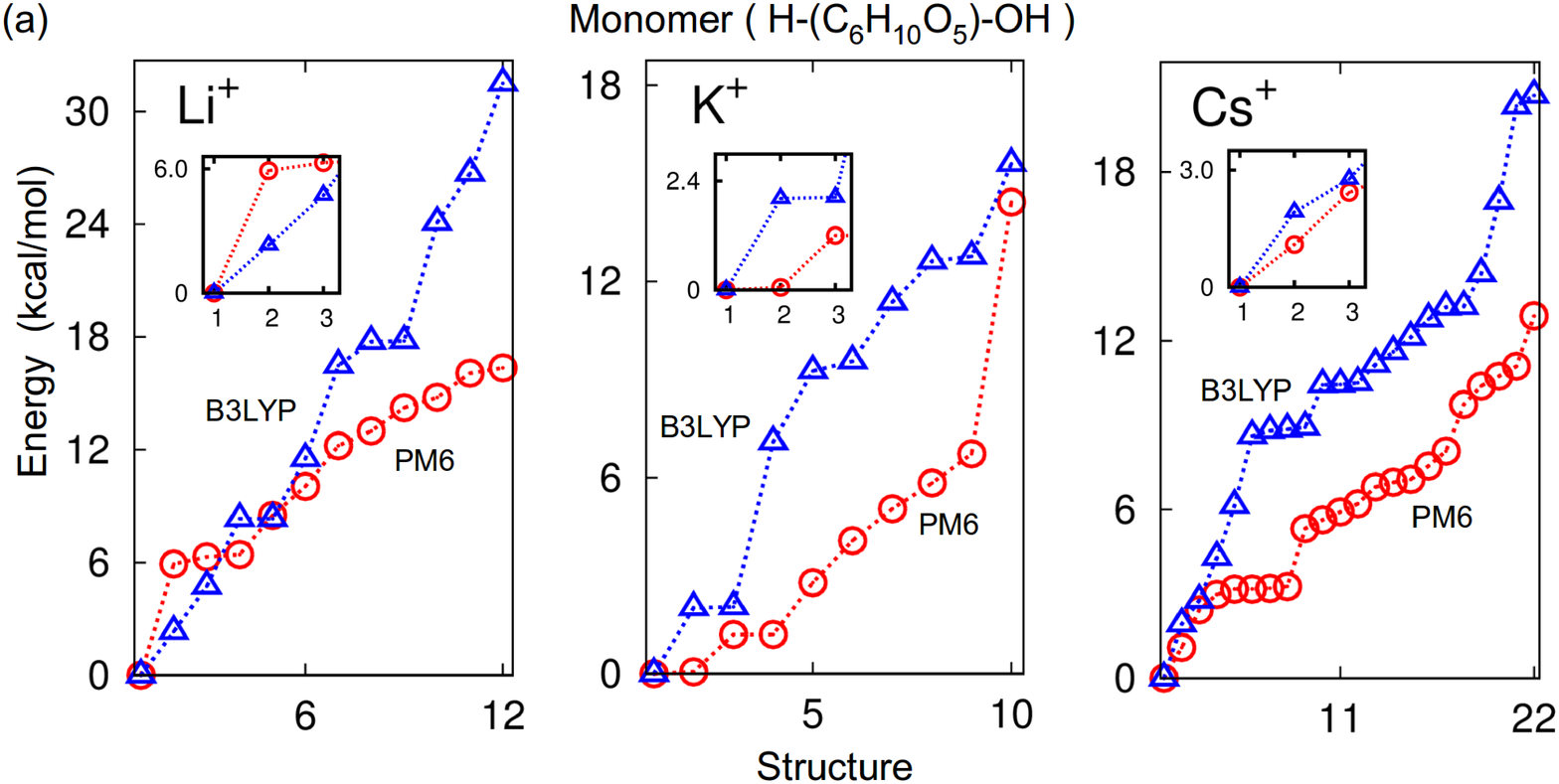}}\vspace{2mm}
\scalebox{0.196}[0.196]{\includegraphics{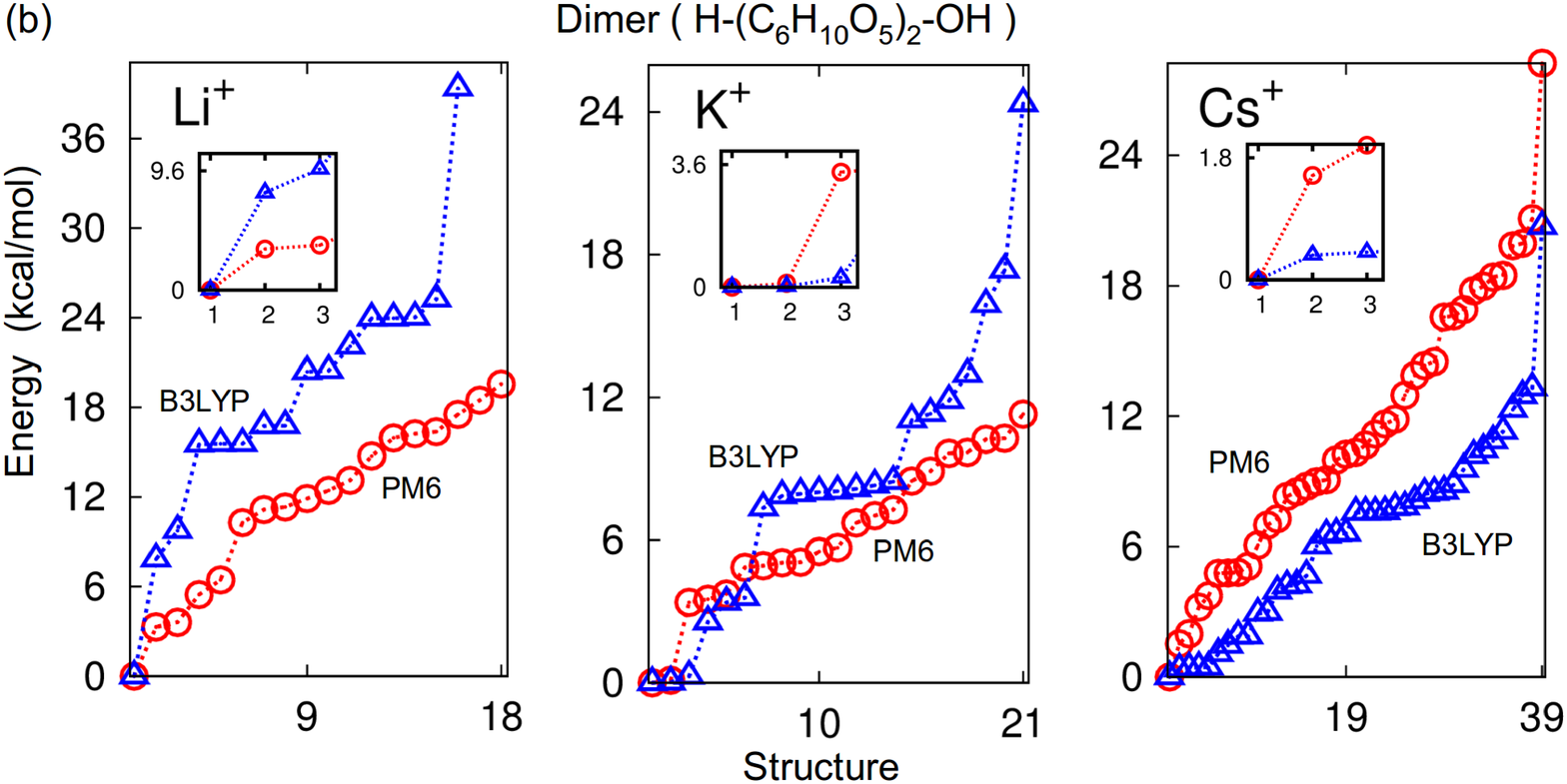}}
\caption{\label{fig:energy}
Energy differences between the stable structures of (a) cellulose
 monomers and (b) cellulose dimers with a cation, discovered by the
 multicomponent artificial force induced reaction (MC-AFIR)
 method~\cite{Maeda;Morokuma:2010,Maeda;Morokuma:2011} with 
 PM6 (red circles) and reoptimized by DFT calculations 
 at the B3LYP/LanL2DZ level of 
 theory (blue triangles). The base point of energy is the energy of
 Structure 1 (i.e. the lowest energy). Three kinds of cations, $\mbox{Li}^{+}$,
 $\mbox{K}^{+}$, and $\mbox{Cs}^{+}$, are considered. The insets show the
 three lowest energies.} 
\end{figure}

\section{Results and discussion}
First, we study the relative stability of the structural data in the
two-step search. 
Figure \ref{fig:energy} shows the energy difference between the stable 
structures discovered by the MC-AFIR method with PM6 (red
circles) and reoptimized by DFT calculations with B3LYP/LanL2DZ (blue
triangles). 
The horizontal axes indicate the structure identification numbers,
while the vertical axes show the energy difference between different
structures. 
The insets show the data of the three lowest energies. 
We obtain cation binding structures with a wide range of
energy. 
Let us focus on the reoptimization results (blue triangles). 
We find that in cellulose dimers multiple lowest-energy
structures appear when the atomic number of cations increases (i.e. the
radius of cations becomes large). 
As for $\mbox{K}^{+}$ [middle panel of Figure \ref{fig:energy}(b)], we
have the two distinct structures within $0.6\,\mbox{kcal}/\mbox{mol}$. 
We count the number of distinct stable structures in the manner of
checking identical inter-atomic distances with threshold distance
$0.1\,\mbox{\AA}$. 
Similarly, as for $\mbox{Cs}^{+}$ [right panel of Figure
\ref{fig:energy}(b)], we have the three distinct lowest-energy
structures. 
Otherwise, the lowest-energy structures are well separated from the
higher-energy ones. 

\begin{figure*}[tbp]
\centering
\scalebox{0.35}[0.35]{\includegraphics{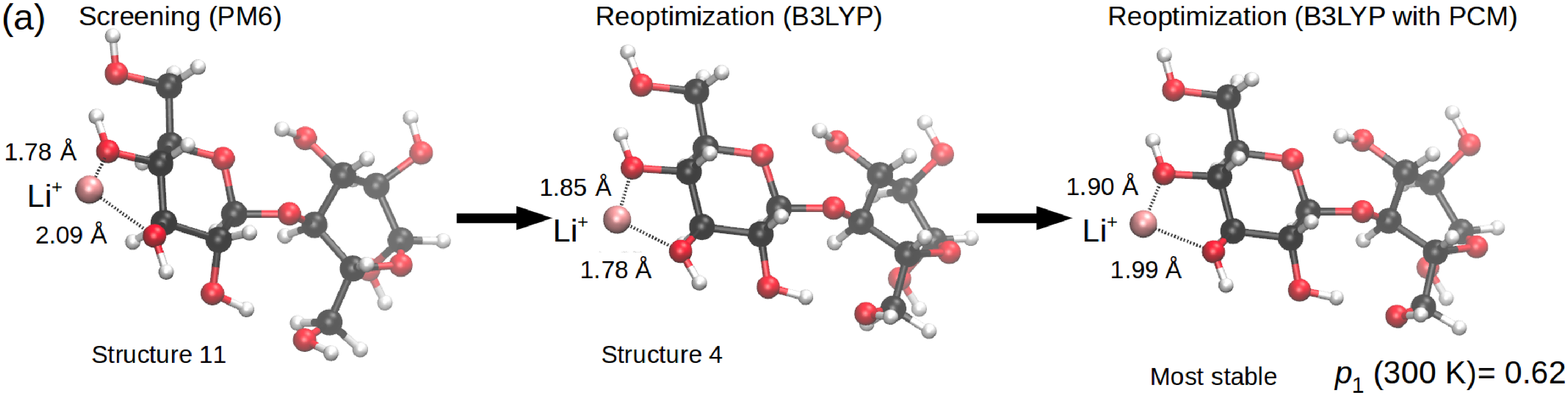}}
\scalebox{0.35}[0.35]{\includegraphics{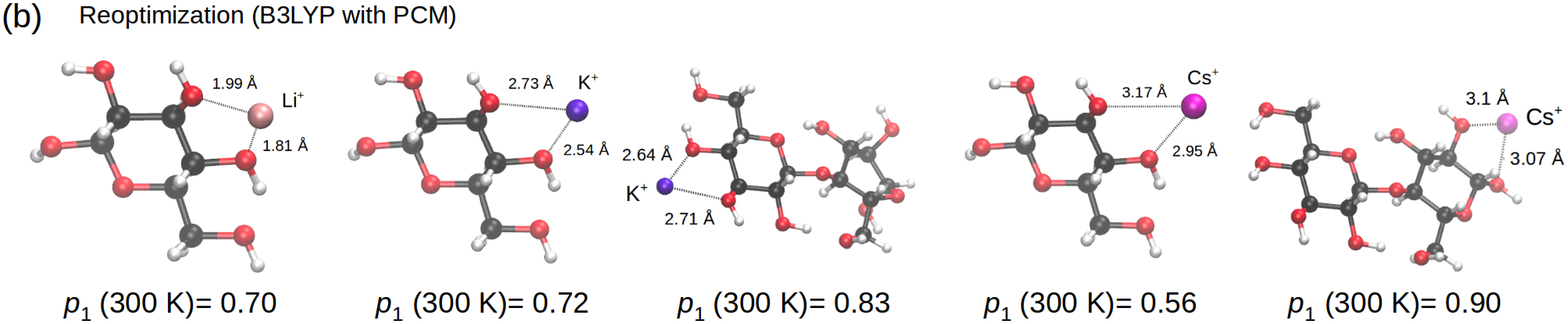}}
\caption{\label{fig:stable_structures}
(a) Sequence of stable structures on cellulose dimers binding
 $\mbox{Li}^{+}$.   
The most stable structure obtained by DFT calculations with PCM
 (right panel) comes from the 11th stable 
 structure (left panel) in the screening performed by the MC-AFIR method
 with PM6, via the 4th stable one (middle panel) obtained by DFT
 calculations without PCM (i.e. gaseous phase). 
The identiciation numbers on the left and middle panels
 correspond to those in the horizontal axes of Figure \ref{fig:energy}. 
On the bottom of the right panel, the probability of finding the most
 stable structure according to partition function
 (\ref{eq:partition_function_structures}) $p_{1}$ at $300\,\mbox{K}$ is
 shown. 
(b) Most stable structures and finding probablity $p_{1}$ at
 $300\,\mbox{K}$ obtained by DFT calculations with PCM, on cellulose
 monomers and dimers binding a cation. The result of a cellulose dimer
 with $\mbox{Li}^{+}$ is shown in the right panel of (a).} 
\end{figure*}

Next, we show the cation binding structures of cellulose molecules. 
Figure \ref{fig:stable_structures} shows the stable structures with
cation-oxygen distances, mainly focusing on the most stable structures
in DFT calculations with PCM. 
All the analyses of molecular structures and visualization were performed
in VMD~\cite{VMD:1996}. 
On the bottom of each panel we show the 
probability of finding the corresponding structure 
according to the canonical ensemble described by partition function
(\ref{eq:partition_function_structures}). 

In Figure \ref{fig:stable_structures}(a), we show a typical sequence of
structures from screening to reoptimization on cellulose dimers with
$\mbox{Li}^{+}$, as well as the most stable structure in DFT
calculations with PCM (right panel). 
The identification numbers on the bottom of the left and
middle panels correspond to those in the horizontal axes of
Figure \ref{fig:energy}. 
The most stable structure in DFT calculations with PCM comes from the
11th stable structure discovered by the MC-AFIR method with PM6.  
Thus, the reoptimization by DFT calculations can alter the
stability order of the structures predicted by semiempirical
methods in screening.  
We stress, however, that the use of GRRM with MOPAC leads to different
kinds of the initial guesses for subsequent high-level
calculations in an unbiased and automated way, rather than the precise
data on molecular systems. 

\begin{figure}[btp]
\centering
\scalebox{0.19}[0.19]{\includegraphics{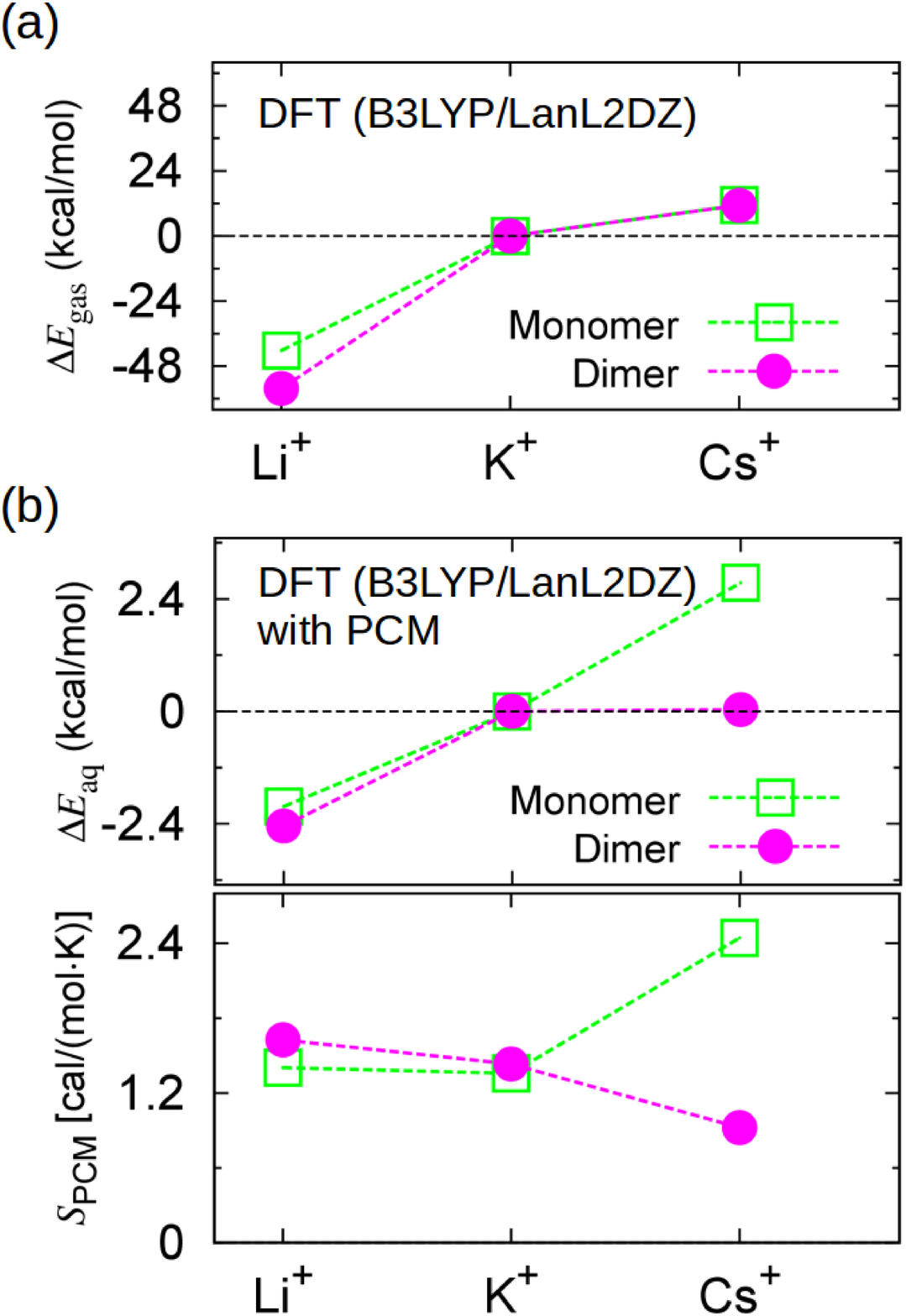}}
\caption{\label{fig:relative_stability}
Energy difference associated with reaction process (\ref{eq:reaction_process}),
 (a) in the gaseous phase and (b) in aqueous solution. 
The energy differences in the gaseous phase contain the DFT-calculation
 energy of the most stable structure, whereas in aqueous
 solution they have the free energy relevant to the canonical ensemble
 of stable structures. 
The evaluation way is explained in the main text
 (Section \ref{sec:methods}). 
A negative value indicates that a cation is more strongly bound to
 a cellulose molecule than $\mbox{K}^{+}$. 
Moreover, on the lower panel of (b), the entropy of stable structures
 according to partition function
 (\ref{eq:partition_function_structures}) is shown.}
\end{figure}

We turn into the probability of finding the most stable structure
in aqueous solutions in Figure \ref{fig:stable_structures}. 
The aqueous solutions described by PCM can change the relative stability
between stable structures in the gaseous phase. 
We find drastic changes for $\mbox{Cs}^{+}$ . 
On cellulose monomers binding $\mbox{Cs}^{+}$, the probability
is less than $0.60$. 
Thus, solvents lead to a broad distribution of stable structures,
although in the gaseous phase a single lowest-energy structure is well
separated from higher-energy ones [right panel of Figure
\ref{fig:energy}(a)]. 
On the other hand, the cellulose dimers binding $\mbox{Cs}^{+}$ have a
single lowest-energy structure with very high probability ($p_{1}=0.9$).
This result is in contrast to the gaseous-phase ones in Figure
\ref{fig:energy}, where there are multiple
lowest-energy structures within $0.6\,\mbox{kcal}/\mbox{mol}$. 
The results for entropy [lower panel of Figure
\ref{fig:relative_stability}(b)] also indicate these effects. 

\begin{table*}[htbp]
\centering
\caption{\label{tab:stat_info}
Statistical information in a search of cation binding
 to cellulose monomers by GRRM14 with PM3, depending on the 
 computational packages of force calculations. 
The search is stopped when the number of equilibrium structures reaches
 $10$. 
The number of force calculations 
($n_{\rm force}$) and the total elapsed time ($t_{\rm elapsed}$) are
 measured.
Then, the mean calculation time of force is evaluated by
$\bar{t}_{\rm force}=t_{\rm elapsed}/n_{\rm force}$. 
Since in GAMESS the PM3 prameter set of $\mbox{Cs}^{+}$ is absent, the
 entries are empty. 
For reference, a force-calcualtion time of DFT calculations at the
 B3LYP/LanL2DZ level of theory is also shown in the last column of the
 second row. 
The structural data in the single-point DFT calculations is built up by 
adding a single cation to the optimized structure of a cellulose
 monomer at the B3LYP/LanL2DZ level of theory. 
The distance between a cation and the center of mass of a monomer is set
 as about $5\,\mbox{\AA}$. 
} 
\begin{tabular}{cccccc}
\hline
Packages of& 
Cations &
$n_{\rm force}$ &
$t_{\rm elapsed}\,(\mbox{sec.})$ &
$\bar{t}_{\rm force}\,(\mbox{sec.})$ &
$t_{\rm force}({\rm DFT})\,(\mbox{sec.})$ \\
force calculations&
&
&
&
&
\\
\hline
 & 
$\mbox{Li}^{+}$&
$7406$ & 
$670$ &
$0.09$ &
--\\
MOPAC & 
$\mbox{K}^{+}$&
$7121$ & 
$610$ &
$0.09$ &
--\\
 & 
$\mbox{Cs}^{+}$&
$5055$ & 
$430$ &
$0.09$ &
--\\
 &
$\mbox{Li}^{+}$&
$5501$ & 
$5261$ & 
$0.96$ &
$135.2$\\
GAMESS &
$\mbox{K}^{+}$&
$4473$ & 
$3880$ & 
$0.87$ &
$151.8$\\
 &
$\mbox{Cs}^{+}$&
-- & 
-- & 
-- &
$140.9$\\
\hline
\end{tabular} 
\end{table*}

Now, we study the amount of cation adsorption on cellulose molecules,
according to reaction process (\ref{eq:reaction_process}). 
Figure \ref{fig:relative_stability} shows the energy differences, 
$\Delta E_{\rm gas}$ and $\Delta E_{\rm aq}$, between the binding of
different cations. 
On the lower panel of Figure \ref{fig:relative_stability}(b), the
entropy of stable structures is also shown.  
In both the gaseous-phase [Figure
\ref{fig:relative_stability}(a)] and aqueous-solution results [Figure
\ref{fig:relative_stability}(b)] cellulose molecules favor binding
$\mbox{Li}^{+}$. 
Moreover, the energy difference increases monotonically with the radius
of the cation. 
However, the variation range of $\Delta E_{\rm aq}$ significantly
narrows, compared to that of the gaseous-phase results. 
Thus, the aqueous solutions described by PCM lead to a reduction in the
relative energy costs associated with cation binding to cellulose. 
This reduction comes purely from the change in electrostatic energy of
molecules since the contributions from the entropy
are quite small as seen in Figure \ref{fig:relative_stability}(b). 
Thus, the two-step structural search is useful for studying 
the distribution of stable structures in molecules. 

Now, we show the efficiency of our approach. 
The computational costs are evaluated on a desktop machine with
the Intel$^{\mbox{\textregistered}}$ Xeon$^{\mbox{\textregistered}}$
E5645 processor, compared to the use of GAMESS. 
We employ the PM3 model~\cite{Stewart:1989,Stewart:1989a}. 
In the screening processes done by GRRM, the number of
force calculations, $n_{\rm force}$, and the total elapsed time of search,
$t_{\rm elapsed}$, are measured. 
Then, the mean time of force calculations, 
$\bar{t}_{\rm force} = t_{\rm elapsed}/n_{\rm force}$, is estimated. 
The screening processes are stopped when the number of discovered
equilibrium structures reaches $10$. 
Table \ref{tab:stat_info} shows the statistical information in a search
of $\mbox{Li}^{+}$ binding to cellulose monomers, by the MC-AFIR method
with PM3. 
Since in GAMESS the PM3 parameter set of $\mbox{Cs}^{+}$ is absent, the
entries are empty.  
For reference, a single-point computational time of force calculations
by DFT via GAMESS is also shown in the last column. 
The table indicates that in a search by GRRM a use of MOPAC
is much more efficient than that of GAMESS. 
In addition, we find that a GRRM search with MOPAC via the interface
program is about $10^{3}$ times faster than the use of DFT force
calculations if the number of force calculations are common to
the two force-calculation methods. 
Thus, our interface program is useful for automatically producing 
various stable structures in large-scale molecules.

Finally, we discuss a range of the applications of the interface
program. 
The validity of our approach depends on that of semiempirical methods. 
Therefore, the use of the interface program is not suitable for
discovering the stable structures of molecular systems with transition
metals and searching chemical reaction pathways including the
rearrangement of covalent bonds. 
The transition-state search should be avoided, as well. 
In contrast, a stable-structure search in organic products is a good
target for our interface program. 
A development of semiempirical methods would extend the application
range. 

\section{Conclusion}
We constructed an interface program between GRRM and MOPAC, to implement
an automated search of stable structures in large-scale molecules with
semiempirical methods. 
We applied this program to studying cation binding to cellulose monomers
and dimers. 
Our approach is a two-step way of discovering stable structures. 
After a search of stable structures by GRRM with PM6, the
resultant structures were reoptimized at the B3LYP/LanL2DZ level of
theory by DFT. 
We found the cation binding structures with a wide range of energy. 
We also demonstrated a way of estimating experimental data on ion adsorption
using well-ordered data of different structures in aqueous
solutions within the PCM of water. 
Moreover, the efficiency of a GRRM search with the interface program
was shown, compared to the use of GAMESS to calculate forces with
semiempirical methods. 
The use of GRRM with MOPAC leads to different kinds of the initial
guesses for high-level calculations in an unbiased and
automated way. 
The present interface program is applicable to various
chemical-reaction-search issues in large-scale molecules, within the
validity of semiempirical methods.

\section*{Acknowledgments}
We would like to thank S. Maeda for his helpful comments.
M.M. acknowledges fruitful discussion with T. Doi and her
collaborators. 
This work is partially supported by Sector of Fukushima Research and
Development in JAEA. 
We thank M. Yui and his colleagues for their support.
The calculations were partially performed on JAEA BX900 supercomputer.
We thank CCSE staff members for their assistance.

\end{document}